# VORTICES IN CIRCUMSTELLAR DISKS

Fred C. Adams and Richard Watkins

*Physics Department, University of Michigan*
*Ann Arbor, MI 48109*

## Abstract

We discuss the physics of vortices in the circumstellar disks associated with young stellar objects. We elucidate the basic physical properties of these localized storm systems. In particular, we consider point vortices, linear vortices, the effects of self-gravity, magnetic fields, and nonlinear aspects of the problem. We find that these vortices can exist in many different forms in the disks of young stellar objects and may play a role in the formation of binary companions and/or giant planets. Vortices may enhance giant planet formation via gravitational instability by allowing dust grains (heavy elements) to settle to the center on a short timescale; the gravitational instability itself is also enhanced because the vortices also create a larger local surface density in the disk. In addition, vortices can enhance energy dissipation in disks and thereby affect disk accretion. Finally, we consider the possibility that vortices of this type exist in molecular clouds and in the disk of the galaxy itself. On all of these size scales, vortices can produce long-lived structures which may correspond to observed structures in these systems.

*Subject headings:* stars: formation – hydrodynamics – planets: formation

## 1. INTRODUCTION

Circumstellar disks play an important role in the star formation process (see, e.g., the reviews of Shu, Adams, & Lizano 1987; Bertout 1989; Beckwith & Sargent 1993). Through many recent studies, we now have a reasonably good understanding of the basic physical properties of these disks, such as total disk mass, temperature distributions, and radial size (see, e.g., Rydgren & Zak 1985; Rucinski 1987; Kenyon & Hartmann 1987; Adams, Lada, & Shu 1987, 1988; Beckwith et al. 1990; Adams, Emerson, & Fuller 1990). Unfortunately, however, many different physical processes occur in these systems and it is not yet known which processes are most important for disk evolution (see, e.g., the review of Adams & Lin 1993). In this paper, we explore a new type of physical process in circumstellar disks. We study the basic physics of vortices, which are essentially long-lived storm systems. We explore possible effects that vortices can have on disk evolution and dynamics. In particular, we show that these vortices can enhance the formation of giant planets within the mechanism of gravitational instability. In addition, vortices can enhance energy dissipation and hence affect disk accretion.

The most well known and well studied example of a fluid vortex in an astrophysical setting is the Great Red Spot of Jupiter (e.g., Ingersoll 1990; see also the special issue of *Chaos* 1994). This stable swirling storm system has lived for many dynamical timescales. Although the exact nature of the Great Red Spot remains somewhat controversial (see, e.g., Marcus 1993 and Petviashvili & Pokhotelov 1992 for different perspectives), the basic physics is known and provides a starting point for this present work. In particular, in this paper we shall focus on quasi-geostrophic vortices, leaving more complex vortex models for future work. Similar vortices arise in most rotating fluid systems, such as the atmosphere and oceans of the Earth (e.g., Hess 1959; Ghil & Childress 1987). We note that vortices have also been studied in the context of galactic disks (Korchagin & Petviashvili 1985, 1991; Korchagin, Petviashvili, & Ryabtsev 1988). These previous studies have not included the self-gravity of the system and have not considered nearly Keplerian rotation curves such as those found in circumstellar disks. We note, however, that vortices in circumstellar disks have been discussed previously in a less modern context by Von Weizsäcker (1944). In addition, two recent papers (Tanga et al. 1995; Barge & Sommeria 1995) also discuss planet formation and dust settling in vortices.

This paper is organized as follows. We first formulate the general problem of a vortex in a self-gravitating circumstellar disk (§2). Our goal is to find analytic (and semi-analytic) solutions for the vortices. We consider several different limiting cases in order to elucidate the basic underlying physics of these storm systems. We consider the case of point vortices in §3; we also present results for linear vortices without self-gravity. Next, we consider higher order effects including self-gravity, nonlinear effects, and magnetic fields (in §4). We discuss applications of vortices to astrophysical problems in §5. We conclude, in §6, with a summary and discussion of our results.

## 2. GENERAL FORMULATION

In this section, we derive the basic equations which describe the physics of self-gravitating vortices in circumstellar disks. In order to obtain analytic results, we make a number of approximations along the way. We consider the vortices to be small compared to the radial position $r$ in the disk; we thus invoke a local approximation. We can then separate the vortical motion from the mean flow of the disk. We introduce both the vorticity and the stream function and obtain equations in terms of these physical quantities.

We begin by writing the fluid equations in a form appropriate for a circumstellar disk. The radial force equation becomes

$$\frac{\partial u}{\partial t} + u \frac{\partial u}{\partial r} + \frac{v}{r} \frac{\partial u}{\partial \phi} - \frac{v^2}{r} + \frac{\partial h}{\partial r} + \frac{\partial \psi}{\partial r} = 0, \qquad (2.1)$$

where $u$ ($v$) is the radial (azimuthal) velocity, $\psi$ is the gravitational potential, and $h$ is the enthalpy. We have thus assumed a barotropic equation of state for the fluid (see, however, §4.5). The azimuthal force equation is

$$\frac{\partial v}{\partial t} + u\frac{\partial v}{\partial r} + \frac{v}{r}\frac{\partial v}{\partial \phi} + \frac{uv}{r} + \frac{1}{r}\frac{\partial}{\partial \phi}\left(\psi + h\right) = 0. \tag{2.2}$$

The continuity equation takes the form

$$\frac{\partial \sigma}{\partial t} + \frac{1}{r}\frac{\partial}{\partial r}(r\sigma u) + \frac{1}{r}\frac{\partial}{\partial \phi}(\sigma v) = 0. \tag{2.3}$$

The gravitational potential $\psi$ obeys the usual Poisson equation, i.e.,

$$\nabla^2 \psi = 4\pi G \rho. \tag{2.4}$$

## 2.1 Local Approximation

We want to consider disturbances which are small compared to the radial extent of the disk. As a general trend, the disturbance size will be roughly comparable to the scale height $H$ of the disk and for thin disks $H/r \ll 1$. We thus work in terms of the variables $x$ and $y$ centered on a point $(r_0, \phi_0)$ rotating with the disk:

$$r = r_0 + x, \tag{2.5}$$

$$y = r(\phi - \phi_0) = r(\phi - \Omega_0 t), \tag{2.6}$$

where $\Omega_0$ is the rotation rate at the center of the vortex (see Figure 1).

In terms of these new variables, the equations of motion become

$$u_t - r\Omega_0 u_y + uu_x + vu_y - \frac{v^2}{r} + \psi_x + h_x = 0, \tag{2.7}$$

$$v_t - r\Omega_0 v_y + uv_x + vv_y + \frac{uv}{r} + \psi_y + h_y = 0, \tag{2.8}$$

$$\sigma_t - r\Omega_0 \sigma_y + \partial_x(u\sigma) + \partial_y(v\sigma) = 0. \tag{2.9}$$

## 2.2 Mean Flow Approximation

As the next simplification, we will separate out the mean flow from the disturbance. In equilibrium, the disk is azimuthally symmetric and has no radial flow, i.e., $u = 0$ and $v = v_0 = r\Omega(r)$. Here, we want to separate the perturbed flow of the vortex from the unperturbed motion; we therefore write the azimuthal velocity $v$ as the sum of parts:

$$v = v_0 + v_1 = r\Omega(r) + v_1, \tag{2.10}$$

where $\Omega(r)$ is the unperturbed rotational speed as a function of radius. This separation of the mean flow can be done in a variety of ways and can lead to somewhat different-looking equations of motion (see, e.g., Petviashvili & Pokhotelov 1992; Ghil & Childress 1987).

Next we define an operator $\mathcal{D}$,

$$\mathcal{D} \equiv \partial_t + [v_0 - r\Omega_0]\partial_y = \partial_t + [r\Omega(r) - r\Omega_0]\partial_y \,. \tag{2.11}$$

The equations of motion now become

$$\mathcal{D}u + uu_x + v_1 u_y - 2\Omega v_1 - \frac{v_1^2}{r} + h_{1x} + \psi_{1x} = 0 \,, \tag{2.12}$$

$$\mathcal{D}v_1 + uv_{1x} + v_1 v_{1y} + u\Omega + \frac{uv_1}{r} + uv_{0x} + h_{1y} + \psi_{1y} = 0 \,, \tag{2.13}$$

$$\mathcal{D}\sigma + \partial_x(u\sigma) + \partial_y(v_1\sigma) = 0 \,. \tag{2.14}$$

In obtaining this form for the force equations, we have used the fact that the mean (unperturbed) flow is determined by the equation

$$\frac{v_0^2}{r} = r\Omega^2 = \partial_x(\psi_0 + h_0) \,. \tag{2.15}$$

We have thus separated out the unperturbed part $h_0$ and the remaining part $h_1$ of the enthalpy function. We have also invoked this same separation for the gravitational potential, i.e., $\psi = \psi_0 + \psi_1$. Although we have separated out the perturbed quantities from the background, notice that we have *not* assumed that the perturbations are small. Thus far, we have retained the full nonlinearity of the equations of motion.

Next, we can combine the two components of the force equation into a single vector equation. Keeping only the leading order terms in $x/r$, we thus obtain

$$\mathcal{D}\mathbf{v} + (\mathbf{v} \cdot \nabla)\mathbf{v} + 2\Omega(\hat{z} \times \mathbf{v}) + \nabla h_1 + \nabla \psi_1 + ur(\partial_x \Omega)\hat{y} = 0 \,, \tag{2.16}$$

where $\mathbf{v} = (u, v_1)$ is the perturbed velocity field. Notice the presence of the Coriolis term $2\Omega(\hat{z} \times \mathbf{v})$, which arises because we are working in a rotating frame of reference.

We also want to write the continuity equation in terms of enthalpy instead of in terms of the surface density $\sigma$. The enthalpy is defined by

$$\nabla h = \frac{a^2}{\sigma} \nabla \sigma \,, \tag{2.17}$$

where $a$ is the effective sound speed of the fluid. We can thus write the continuity equation in the form

$$\mathcal{D}h_1 + \mathbf{v} \cdot \nabla h + a^2 \nabla \cdot \mathbf{v} = 0 \,, \tag{2.18}$$

where we have used the fact that the unperturbed state is independent of both time $t$ and the variable $y$ so that $\mathcal{D}h = \mathcal{D}h_1$.

*2.3 Vorticity*

Next, we want to write the equations of motion in terms of vorticity, instead of in terms of velocity. This approach makes sense because we are interested in vortex solutions and vorticity is the relevant physical quantity for vortices. The vorticity $\omega$ is defined by

$$\omega \equiv \hat{z} \cdot (\nabla \times \mathbf{v}) \,, \tag{2.19}$$

i.e., we consider only the $\hat{z}$ component of the vorticity.

We now take the curl of the force equation (in the form of equation [2.16]). Next, we combine the continuity equation [2.18] with this newly derived vorticity equation. After some rearrangement, this equation of motion becomes

$$\mathbf{D}\left[\frac{\omega + 2\Omega + r\partial_x \Omega}{\sigma}\right] = 0, \qquad (2.20)$$

where we have defined the operator $\mathbf{D}$ to be the total advective derivative operator:

$$\mathbf{D} \equiv \mathcal{D} + \mathbf{v} \cdot \nabla = \partial_t + r(\Omega - \Omega_0)\partial_y + \mathbf{v} \cdot \nabla. \qquad (2.21)$$

### 2.4 Geostrophic Balance

We next use the fact that the fluid can be in some kind of generalized geostrophic balance. Geostrophic balance occurs when pressure forces are balanced by Coriolis forces (so that the inertial forces are negligible). In this work, we generalize this concept to include gravity, so that both pressure and gravitational forces are balanced by the Coriolis force (see Figure 2). In particular, we write the vector velocity $\mathbf{v}$ in the form

$$\mathbf{v} = \frac{1}{2\Omega_0}\left[\hat{z} \times (\nabla h_1 + \nabla \psi_1)\right], \qquad (2.22)$$

where $\Omega_0$ is the value of the rotation rate in the disk at the location of the origin of the local $(x, y)$ coordinate system. In the geostrophic approximation, the vorticity can be written

$$\omega = \frac{1}{2\Omega_0}\left(\nabla^2 h_1 + \nabla^2 \psi_1\right). \qquad (2.23)$$

Thus, our equation of motion [2.20] now becomes

$$\mathbf{D}\left[\frac{1}{2\Omega_0 \sigma}\left(\nabla^2 h_1 + 4\pi G \rho_1\right) + \frac{2\Omega + r\partial_x \Omega}{\sigma}\right] = 0, \qquad (2.24)$$

where we have used the first order part of the Poisson equation [2.4] to eliminate the gravitational potential.

The equation of motion [2.24] represents the basic model for vortices in this paper. In analogous work in other fields, this model is usually called the quasi-geostrophic (QG) approximation (e.g., Ghil & Childress 1987). This model can be generalized in several different ways (e.g., Petviashvili & Pokhotelov 1992; Sutyrin 1994); for example, additional correction terms can be included in the geostrophic balance equation [2.22]. However, in this paper, we want to elucidate the basic physical properties of vortices in circumstellar disks. We thus focus our discussion on the simple model implied by equation [2.24]; further generalizations of this model are left for future work.

## 2.5 Summary of Approximations

The equation of motion [2.24] describes vortices in a circumstellar disk with self-gravity included. We now define a quantity $\Upsilon$, which is a generalized vorticity per unit surface density,

$$\Upsilon \equiv \frac{1}{2\Omega_0 \sigma} \left( \nabla^2 h_1 + 4\pi G \rho_1 \right) + \frac{2\Omega + r\partial_x \Omega}{\sigma} \,. \tag{2.25}$$

We refer to this quantity $\Upsilon$ as the *generalized vortensity*. The first part of equation [2.25] represents the vortensity due to a perturbation; the second part represents the vortensity of the unperturbed disk flow. With this definition, our equation of motion can be written in the seemingly simple form

$$\mathbf{D}\Upsilon = 0 \,. \tag{2.26}$$

This equation thus implies that the total generalized vortensity is advectively conserved. Vortensity can, in principle, be freely transferred between the mean flow and the perturbation, but the total vortensity is conserved.

Before considering the solutions to the equation of motion [2.26], we summarize the approximations we have made to obtain this form. In particular, we have made the local approximation and the assumption of quasi-geostrophic balance.

The local approximation assumes that the vortex size $\Lambda$ is small compared to the radius $r_0$ of the disk (at the vortex center). Since we are working with two dimensional thin disks, we also require that the vortex is large compared to the disk scale height $H$. Our treatment thus implies the following ordering of length scales:

$$H \ll \Lambda \ll r_0 \,. \tag{2.27}$$

The approximation of quasi-geostrophic balance assumes that the most important part of the perturbed velocity field corresponds to vortical motions about the vortex center (the origin of local coordinates in our treatment). To see what this approximation implies mathematically, consider the force equation [2.16]. The third, fourth, and fifth terms of this equation give us the geostrophic approximation [2.22]. Thus, for consistency, we must require the ordering

$$\frac{\langle v \rangle}{\tau} \sim \frac{\langle v \rangle^2}{\Lambda} \ll \Omega \langle v \rangle \sim \frac{h_1}{\Lambda} \,, \tag{2.28}$$

where $\langle v \rangle$ is an average perturation velocity and $\tau$ is the timescale over which the velocity changes. This ordering requirement can be satisfied provided that (1) the timescale for the vortex to change its velocity profile is long ($\tau$ large), (2) the mean perturbation speed is subsonic (so that $\langle v \rangle^2 < h_1$), and (3) the mean vortex rotation rate is slower than the local rotation speed of the disk (so that $\langle v \rangle / \Lambda < \Omega$).

## 3. RESULTS FOR POINT VORTICES AND LINEAR VORTICES

In this section, we consider the simplest type of vortices in circumstellar disks. We discuss the limiting case of point vortices in §3.1 and general linear vortices without self-gravity in §3.2.

## 3.1 Point Vortices

The simplest type of vortex is a point vortex, i.e., a vortex with vorticity localized at a point in space. Vortices of this type are the solutions to the simplest form of our equation of motion. For example, if we consider the rotation rate $\Omega$ and the total surface density $\sigma$ of the disk to both be constant, then the equation of motion [2.26] simplifies to the form

$$\mathbf{D}\omega = 0 \,. \tag{3.1}$$

In other words, in this limiting case, the vorticity itself is advectively conserved. This equation of motion has solutions which correspond to point vortices. Mathematically, we can write these solutions as

$$\omega = \Gamma \delta^2(\mathbf{x} - \mathbf{x}_0) \,, \tag{3.2}$$

where $\delta^2(\mathbf{x})$ is the two-dimensional Dirac Delta function and $\mathbf{x}_0$ is the position of the vortex. The quantity $\Gamma$, which determines the magnitude of the vorticity, is usually called the circulation.

Once the vorticity solution is specified in the form of equation [3.2], we can determine the physical structure of the vortex (Marcus 1993). We find the velocity field in a manner completely analogous to finding the magnetic field produced by a line-like current (e.g., Jackson 1975). This calculation implies

$$\mathbf{v} = \frac{\Gamma}{2\pi\varpi}\,\hat{\varphi} \,, \tag{3.3}$$

where we have introduced local cylindrical coordinates centered on the position of the point vortex ($\varpi$ is the local radial coordinate). The density enhancement is determined by the condition [2.22] of geostrophic balance. Notice that positive density enhancements correspond to negative values of the circulation $\Gamma$. In other words, the vortex motion must be clockwise (in the $-\hat{\varphi}$ direction) for a positive density vortex. For this case, the enthalpy perturbation takes the form

$$h_1 = \frac{\Omega_0 |\Gamma|}{\pi} \log[\varpi_0/\varpi] \,, \tag{3.4}$$

where $\varpi_0$ is an integration constant which determines the physical size of the vortex (see below). Finally, we can write the vortex solution in terms of the surface density via

$$\frac{\sigma_1}{\sigma_0} = \left(\frac{\varpi_0}{\varpi}\right)^\gamma \,, \tag{3.5}$$

where the index $\gamma$ is defined by

$$\gamma \equiv \frac{\Omega_0 |\Gamma|}{\pi a^2} \,. \tag{3.6}$$

Notice that there exists a contradiction in the above argument. We have assumed that the total surface density is a constant in order to obtain the simplified equation of motion [3.1]. On the other hand, our solution implies a surface density perturbation [3.4 – 3.6].

It turns out that this solution is a limiting case of the more general problem considered in the next subsection. This present solution is valid in the limit that the parameter $k_R \to 0$, where $k_R$ is the Rossby wavenumber defined below (equation [3.10]).

*3.2 Linear Vortices Without Self-Gravity*

For the next higher order of approximation, we consider the case of vortices with no self-gravity and we consider only the leading order perturbations for $\Omega$ and $\sigma$. We also specialize to the case of a purely Keplerian rotation curve, i.e., $\Omega \sim r^{-3/2}$ and $r\partial_x \Omega = -(3/2)\Omega$. In this approximation, we introduce a stream function $\Phi$ defined by

$$\Phi = \frac{h_1}{2\Omega_0}. \tag{3.7}$$

Next, we separate the total surface density $\sigma$ into the unperturbed part $\sigma_0(r)$ and the perturbation $\sigma_1$. Furthermore, the unperturbed surface density itself can be expanded as

$$\sigma_0(x) = \sigma_0(r) + \frac{d\sigma_0}{dr}\bigg|_0 x = \sigma_0(r)\big[1 - px/r\big], \tag{3.8}$$

where we have assumed that the unperturbed surface density has a power-law distribution with index $p$. Recall that $r$ is the radial coordinate of the vortex center. To leading order, the generalized vortensity can be written

$$\Upsilon = \frac{1}{\sigma_0(r)}\left[\nabla^2\Phi + \frac{1}{2}(\Omega - \Omega_0) - \frac{\Omega_0^2}{a^2}\Phi + \Omega_0\frac{px}{2r}\right], \tag{3.9}$$

where we have subtracted out the constant contribution, $\sim \Omega_0/\sigma_0$, of the background state. Keep in mind that we have assumed that the perturbed surface density is small, i.e., $|\sigma_1| \ll \sigma_0$. In §4.2, we show how to generalize this result.

Departing slightly from the usual conventions (e.g., see Marcus 1993), we define a Rossby wavenumber $k_R$, i.e.,

$$k_R = \frac{\Omega_0}{a}. \tag{3.10}$$

For circumstellar disks, this Rossby wavenumber is the inverse of the disk scale height ($k_R = 1/H$). We can also expand the term $\Omega - \Omega_0$ to leading order in $x = r - r_0$. We thus obtain the expression

$$\Upsilon = \frac{1}{\sigma_0(r)}\left[\nabla^2\Phi - k_R^2\Phi + \alpha_0 x\right], \tag{3.11}$$

where we have defined a parameter $\alpha_0$,

$$\alpha_0 = \Omega_0(2p - 3)/4r. \tag{3.12}$$

Notice that in general $\alpha_0$ can be either positive or negative, depending on whether the surface density $\sigma_0(r)$ decreases faster or slower than the rotation curve $\Omega(r)$. For the particular case of $p = 3/2$, we obtain $\alpha_0 = 0$. Theories of the formation of protostellar disks suggest that disk density profiles will have indices in the range $1 - 2$ (see, e.g., Cassen & Moosman 1981; Terebey, Shu, & Cassen 1984). Thus, we expect $\alpha_0$ to be small and we can take $p \approx 3/2$ and hence $\alpha_0 \approx 0$ as a starting approximation.

The solutions for the stream function, the velocity field, and the surface density perturbation are qualitatively similar to those of the point vortices discussed in the previous subsection. We can find these solutions as follows. We first construct generalized vortensity profiles $\Upsilon$ that satisfy the conservation condition of equation [2.26]. We can then invert the differential operator appearing in equation [3.11] to obtain the stream function $\Phi$. Given the stream function, we can determine the velocity field of the vortex through the relation

$$\mathbf{v} = \hat{z} \times \nabla \Phi. \tag{3.13}$$

For example, suppose we take a solution $\Upsilon(\mathbf{x})$ for the vortensity, which could be a collection of point vortices, or perhaps a constant over some area. We thus obtain the solution for the stream function in the form

$$\Phi(\mathbf{x}) = \sigma_0(r) \int d^2\mathbf{x}' G(\mathbf{x}, \mathbf{x}') \Upsilon(\mathbf{x}'), \tag{3.14}$$

where we have taken $\alpha_0 = 0$ and where $G(\mathbf{x}, \mathbf{x}')$ is the Green's function corresponding to the differential operator in equation [3.11]. This operator is simply the modified Helmholtz operator (in two dimensions). The appropriate Green's function is made up of modified Bessel functions (e.g., Arfken 1985) and can be written in the form

$$G(\mathbf{x}, \mathbf{x}') = \frac{1}{2\pi} K_0(k_R |\mathbf{x} - \mathbf{x}'|), \tag{3.15}$$

where $K_0$ is the modified Bessel function of order zero. Note that this is the appropriate Green's function for the boundary condition that $G \to 0$ as $|\mathbf{x} - \mathbf{x}'| \to \infty$. Notice also that the Rossby wavenumber $k_R$ determines the length scale on which the stream function $\Phi$ changes; the velocity field $\mathbf{v}$ changes on this same length scale. The lengthscale $1/k_R$ is comparable to the scale height $H$ in the disk; thus, the ordering requirement of equation [2.27] can be satisfied, but only by a small margin.

Thus, for a given solution $\Upsilon(\mathbf{x})$ for the vortensity, we can write the solution for the stream function in the form

$$\Phi/\sigma_0 = \frac{1}{2\pi} \int d^2\mathbf{x}' K_0(k_R |\mathbf{x} - \mathbf{x}'|) \Upsilon(\mathbf{x}'). \tag{3.16}$$

For the purpose of illustration, let the generalized vortensity be that of a point vortex, i.e.,

$$\Upsilon = \frac{\Gamma}{\sigma_0} \delta^2(\mathbf{x}). \tag{3.17}$$

For a brief discussion of more general solutions for the vortensity $\Upsilon$, see Appendix A. With this choice [3.17] for $\Upsilon$, the stream function becomes

$$\Phi = \frac{\Gamma}{2\pi} K_0(k_R \varpi) = \frac{h_1}{2\Omega_0}, \tag{3.18}$$

where $\varpi$ is the local radial coordinate centered on the vortex. Equation [3.18] thus determines the density perturbation. The velocity field of the vortex produced by the stream function can be determined from equation [3.13], which implies that

$$\mathbf{v} = \frac{\Gamma}{2\pi}\left[\hat{z} \times \nabla K_0(k_R\varpi)\right] = \frac{\Gamma k_R}{2\pi}\left[-K_1(k_R\varpi)\right]\hat{\varphi}, \qquad (3.19)$$

where we have used the relation $K_0'(z) = -K_1(z)$ in obtaining the second equality. This solution is thus similar to the point vortex solution found in the previous subsection. In the limit of small local radius, $\varpi \to 0$, we find that $\mathbf{v} \propto \Gamma/\varpi$ (compare with equation [3.3]). However, in the opposite limit $\varpi \to \infty$, we find $\mathbf{v} \propto (\Gamma k_R)z^{-1/2}\,\mathrm{e}^{-z}$, where $z = k_R\varpi$. The perturbation is thus exponentially suppressed at large radial distances from the vortex center, i.e., for $k_R\varpi \gg 1$.

Next, we consider vortices which have a constant vortensity over some area (and zero vortensity outside). Whenever the shear flow can be neglected (so that we can take $\Omega = \Omega_0$), we obtain circularly symmetric vortices. Here, we consider vortices with vortensity profiles which are step functions in radius, i.e., we consider the form

$$\Upsilon = \frac{\Gamma}{\pi R^2}\,\Theta(R - r), \qquad (3.20)$$

where $\Theta$ is the usual Heaviside step function and where $R$ is the radius of the area with nonzero vortensity. Using this form in equation [3.14], we can directly calculate the stream function, the density perturbations, and the corresponding velocity fields. The results are shown in Figures 3 and 4 for the cases with $k_R R = 1/5$ and 5. The solutions for the velocity field and the vorticity can be found analytically (Marcus 1993). For small radii, $0 \leq \varpi \leq R$, we obtain

$$\mathbf{v} = \frac{\Gamma}{\pi R}\,K_1(k_R R)\,I_1(k_R\varpi)\,\hat{\varphi}, \qquad (3.21)$$

and

$$\omega = \frac{\Gamma k_R}{\pi R}\,K_1(k_R R)I_0(k_R\varpi), \qquad (3.22)$$

whereas for larger radii, $\varpi \geq R$, we obtain

$$\mathbf{v} = \frac{\Gamma}{\pi R}\,I_1(k_R R)\,K_1(k_R\varpi)\,\hat{\varphi}, \qquad (3.23)$$

and

$$\omega = -\frac{\Gamma k_R}{\pi R}\,I_1(k_R R)K_0(k_R\varpi), \qquad (3.24)$$

where $I_\nu$ and $K_\nu$ are modified Bessel functions. In the limit of small $k_R R$, these solutions approach the point vortex solutions discussed above. For $k_R R \gg 1$, the vortices have "quiet centers" and most of the vorticity is concentrated near the boundary.

In the presence of shear flow (neglected here), vortex solutions with constant $\Upsilon$ become noncircular. For the special case of linear shear and $k_R \to 0$, the vortices have the shape of perfect ellipses (Moore & Saffman 1971). For the more general case, the vortices are completely specified by the location of their boundaries, but the boundary shapes must be found numerically (for further discussion, e.g., see Marcus 1993).

## 4. HIGHER ORDER EFFECTS

In this section, we consider higher order effects on the vortices. In particular, we consider the effects of self-gravity (§4.1), nonlinearity (§4.2), magnetic fields (§4.3), and the difference between vortices with positive and negative circulations (§4.4). We also briefly address the important issue of vortex generation in §4.5.

### 4.1 Leading Order Effects of Self-Gravity

In order to understand the effects of self-gravity on vortices, we proceed in the same fashion as in the previous subsection but we include the gravitational term. Repeating the steps leading to equation [3.11], we obtain

$$\Upsilon = \frac{1}{\sigma_0(r)} \left[ \nabla^2 \Phi - k_R^2 \Phi + \frac{2\pi G \rho_1}{\Omega_0} \right], \qquad (4.1)$$

where we have again taken $\alpha_0 = 0$. We now make a heuristic approximation to simplify the gravity term,

$$\rho_1 = \frac{\sigma_1}{Z} = \frac{\sigma_0}{a^2 Z} h_1 = \frac{2\sigma_0 \Omega_0}{a^2 Z} \Phi, \qquad (4.2)$$

where $Z$ is an effective total height of the disk. We also define an effective Jeans wavenumber $k_J$ through the relation

$$k_J^2 \equiv \frac{4\pi G \sigma_0}{a^2 Z}. \qquad (4.3)$$

The expression for vortensity becomes

$$\Upsilon = \frac{1}{\sigma_0(r)} \left[ \nabla^2 \Phi - k_R^2 \Phi + k_J^2 \Phi \right]. \qquad (4.4)$$

We can now directly see the effects of gravity on the structure of the vortex. In the previous example, we saw that the Rossby wavenumber $k_R$ determines the lengthscale on which the vortex structure can change. Gravitational effects appear in the equation with the opposite sign; as a result, gravity makes the effective Rossby wavenumber smaller and hence makes the effective Rossby radius larger.

An important crossover point occurs when the gravitational contribution exceeds the (old) Rossby wavenumber contribution and the total effective Rossby radius becomes imaginary. This crossover occurs when $k_R = k_J$, which implies that

$$4\pi G \sigma_0 = \Omega_0^2 Z. \qquad (4.5)$$

If we also assume that the total effective height $Z$ of the disk is twice the usual thermal scale height in the disk, i.e., $Z \approx 2a/\Omega_0$, we obtain the crossover condition in the form

$$\frac{\Omega_0 a}{\pi G \sigma_0} = Q_T \approx 2, \qquad (4.6)$$

where $Q_T$ is the Toomre stability parameter for gaseous disks (Toomre 1964). Here, when $Q_T$ becomes less than 2, the effective Rossby wavenumber becomes imaginary. Notice that in identifying our crossover condition with $Q_T$, we have assumed that the rotation curve is purely Keplerian so that the epicyclic frequency $\kappa$ is equal to the rotational frequency $\Omega$.

The parameter $Q_T$ must be larger than unity in order for the disk to be stable to axisymmetric disturbances. In fact, consideration of non-axisymmetric perturbations in circumstellar disks suggests that the stability parameter must be (at least) as large as $Q_T \sim 2$; otherwise the disks would be highly unstable and would have short lifetimes (e.g., Adams, Ruden, & Shu 1989; Shu et al. 1990; Tomley, Cassen, & Steiman-Cameron 1991; Adams & Benz 1992; Laughlin 1994). As a result, many disk systems tend to live close to this crossover point. As disks evolve in time, they eventually lose mass to the central star, the value of $Q_T$ increases, and the effective Rossby wavenumber approaches the true Rossby wavenumber.

The effect of self-gravity is thus to make the effective Rossby radius longer. It is useful to define an effective Rossby wavenumber $k_{\text{eff}}$,

$$k_{\text{eff}}^2 \equiv k_R^2 \left\{ 1 - \frac{2}{Q_T} \right\}. \qquad (4.7)$$

The equation of motion which determines the stream function can then be simplified to the form

$$\nabla^2 \Phi - k_{\text{eff}}^2 \Phi = \sigma_0 \Upsilon(\mathbf{x}). \qquad (4.8)$$

Solutions for $\Phi$ can be found in a manner analogous to that presented in the previous section.

For purposes of illustration, we consider the special case in which the effective Rossby wavenumber vanishes. In this particular case, the appropriate Green's function for the differential operator in equation [4.8] (i.e., a two dimensional Laplacian operator) is given by

$$G(\mathbf{x}, \mathbf{x}') = -\frac{1}{2\pi} \log |\mathbf{x} - \mathbf{x}'|. \qquad (4.9)$$

The solutions for the stream function are thus given by

$$\Phi = -\frac{\sigma_0}{2\pi} \int d^2 \mathbf{x}' \Upsilon(\mathbf{x}') \log |\mathbf{x} - \mathbf{x}'|. \qquad (4.10)$$

Next, we consider the simplest case in which the vortensity is localized and can be represented as a delta function as in equation [3.17]. We thus obtain

$$\Phi = -\frac{\Gamma}{2\pi} \log |\mathbf{x}|, \qquad (4.11)$$

$$\mathbf{v} = \frac{\Gamma}{2\pi\varpi}\hat{\varphi}, \qquad (4.12)$$

where $\varpi$ is the local radial coordinate and the velocity is in the local azimuthal direction $\hat{\varphi}$. Thus, the velocity field of the perturbation is the same as that of the point vortices studied in §3.

### 4.2 Nonlinear Density Effects

In this subsection, we generalize our previous results by considering the case when the perturbed surface density $\sigma_1$ is *not* small compared to the unperturbed state. For an isothermal equation of state, the enthalpy $h$ is related to the surface density through the relation

$$h = a^2 \log \sigma, \qquad (4.13)$$

which implies that the perturbed part of the enthalpy can be written in the form

$$h_1 = h - h_0 = a^2 \log\left[\frac{\sigma}{\sigma_0}\right]. \qquad (4.14)$$

Using this result, we can write the generalization of equations [3.9] and [3.11] in the form

$$\nabla^2 \Phi + \frac{1}{2}\Omega + \frac{2\pi G \rho_1}{\Omega_0} = \Upsilon(\mathbf{x})\sigma_0 \exp[2\Omega_0 \Phi/a^2]. \qquad (4.15)$$

In order to isolate the nonlinear effects in this equation, we consider the particular case in which the surface density $\sigma_0$ and the sound speed $a$ are all constant. This approximation is valid when these quantities vary sufficient slowly compared to the stream function. We define a reduced stream function according to

$$\mu \equiv \frac{2\Omega_0}{a^2}\Phi = \frac{h_1}{a^2}. \qquad (4.16a)$$

We also define a non-dimensional variable $\mathbf{z}$ by

$$\mathbf{z} \equiv k_R \mathbf{x} = \frac{\Omega_0}{a}\mathbf{x}. \qquad (4.16b)$$

With these definitions, the equation of motion for the vortensity can be written

$$\begin{aligned}\nabla_z^2 \mu &= \frac{2\Upsilon\sigma_0}{\Omega_0}e^\mu - \frac{4\pi G \sigma_0}{\Omega_0^2 Z}\mu - 1, \\ &= A e^\mu - B\mu - 1.\end{aligned} \qquad (4.17)$$

The quantity $A$ defined above is the ratio of the vortensity in the vortex to that of the unperturbed circumstellar flow. The second constant $B$ is the ratio of the Jeans wavenumber (defined in equation [4.3]) to the Rossby wavenumber (defined in equation [4.4]); thus, the constant $B$ determines the relative importance of self-gravity.

This nonlinear form of the equation of motion can lead to qualitatively new behavior under some circumstances. In the general case (see Appendix A), the vortensity $\Upsilon$ is some function of the stream function ($\Phi$ or $\mu$), so that in equation [4.17], the quantity $A = A(\mu)$. Thus, the right hand side of equation [4.17] can be considered as some function $F(\mu)$. In other words, the equation becomes

$$\nabla_z^2 \, \mu = A(\mu) \, e^\mu - B\mu - 1 \equiv F(\mu) = -\frac{dV}{d\mu}. \tag{4.18}$$

The function $F(\mu)$ thus plays the role of a force and its integral $V(\mu)$ plays the role of a potential. Nonlinear equations of this type are well studied in the context of quantum field theory (e.g., Coleman 1985; Rajaraman 1987). The properties of the function $V(\mu)$ determine the allowed types of solutions for $\mu(\mathbf{z})$. In particular, when $V$ has both a double zero and a simple zero (when considered as a function of the reduced field $\mu$), then *solitary wave* solutions are allowed (see, e.g., Drazin & Johnson 1989; Infeld & Rowlands 1990); these solutions correspond to spatially localized wavelike structures. Unfortunately, however, a complete analysis of all of the possible solutions to the nonlinear equation of motion is beyond the scope of this present work and must be left for the future.

### 4.3 Effects of Magnetic Fields

In this section, we consider the effects of magnetic fields on these vortices. For the sake of definiteness, we will consider a magnetic field which threads the disk in the vertical direction only, i.e.,

$$\mathbf{B} = B(x,y)\hat{z} \, . \tag{4.19}$$

We note that the no-monopole condition, $\nabla \cdot \mathbf{B} = 0$, is automatically satisfied with this type of magnetic field. The field adds a force term to the equations of motion; this force $\mathbf{f}_M$ has the form

$$\mathbf{f}_M = \frac{1}{4\pi\rho}(\nabla \times \mathbf{B}) \times \mathbf{B} = \frac{1}{8\pi\rho} \nabla B^2 \, . \tag{4.20}$$

Thus, one important effect of the magnetic field is to add an additional contribution to the pressure force. For the particular field geometry used here, the magnetic field simply changes the effective equation of state of the fluid. The change is thus quantitative, but not qualitative.

For completeness, we note that the continuity equation for the magnetic field can be written in the form

$$\mathbf{D}\left[\frac{B}{\sigma}\right] = 0 \, . \tag{4.21}$$

In other words, the magnetic field strength per unit surface density is also advectively conserved.

The presence of a magnetic field also allows for a new channel of energy dissipation. In general, if the fluid has a finite electrical resistivity $\eta$, the magnetic field will tend to diffuse according to the equation of motion

$$\frac{\partial \mathbf{B}}{\partial t} + \nabla \times (\mathbf{B} \times \mathbf{u}) = \eta \, \nabla^2 \mathbf{B} \, . \tag{4.22}$$

In the limit of $\eta \to 0$, we recover the usual flux freezing assumption. For a given resistivity $\eta$, the magnetic diffusion timescale $\tau_M$ is given by

$$\tau_M \sim \Lambda^2/\eta \,, \tag{4.23}$$

where $\Lambda$ is the physical lengthscale of the system (i.e., the vortex size). Using typical values for $\eta$ (see Shu 1992), we find that the diffusion timescale is much longer than the expected disk lifetime ($\sim 10^6 - 10^7$ yr, e.g., Strom, Edwards, & Skrutskie 1993).

### 4.4 The Prograde/Retrograde Asymmetry

Within Keplerian circumstellar disks, vortices with positive circulation behave differently than vortices with negative circulation. As we have discussed previously, in the present context, a vortex with a *positive* density perturbation must have a *negative* circulation in order to achieve geostrophic balance. In other words, a vortex with a positive density perturbation must rotate clockwise in the *rotating* frame of reference in which we have been working. Now, suppose a planet (or other secondary body) forms within a vortex. This planet would naively appear to be rotating in a retrograde sense. However, this apparent retrograde rotation (say, at a rate $\Omega_P$) must be corrected for the fact that the frame of reference is rotating as well in the opposite direction (at the local Keplerian rotation rate $\Omega$). As long as the "local" rotation rate $\Omega_P$ is smaller than the Keplerian rotation rate $\Omega$, the secondary body will rotate in a prograde sense in the inertial frame of reference. Notice also that if $\Omega_P \geq \Omega$ (which would lead to a real retrograde rotation of the planet), then the approximations leading to our equations of motion break down.

Another important difference between vortices with positive and negative circulation is their lifetime and stability. In a Keplerian disk, the mean flow velocity decreases with radial distance to the star. In our local frame of reference centered on the vortex, the mean flow velocity is negative for $x > 0$ and positive for $x < 0$; this effect is best illustrated pictorially (see Figure 5). As a result, vortices with negative (clockwise) circulation rotate "with the flow", whereas vortices with positive (anti-clockwise) circulation must directly oppose the flow; these latter vortices are defined as "adverse". Detailed numerical studies (cf. the review of Marcus 1993) show that adverse vortices are quickly ripped apart into long filaments and fragments, i.e., their lifetimes are very short. On the other hand, vortices of the other type tend to merge and grow into coherent structures. A complete discussion of the lifetime and stability of vortices is beyond the scope of this present paper. We stress, however, that the vortices with positive density perturbations (negative circulation) are the ones most likely to live a long time in a Keplerian flow.

### 4.5 Vortex Generation

Perhaps the most important unresolved issue is the manner in which these vortices are generated in the first place. Thus far in this paper, we have found many different solutions to the equations of fluid dynamics; these solutions represent *possible* behavior of the circumstellar disk. However, before the astrophysical importance of vortices can be established, one must understand the generation mechanism. In particular, the timescales for vortices to grow and decay are important. In this section, we (crudely) estimate a timescale for the vortensity (or vorticity) to change in a circumstellar disk.

One can argue that all differentially rotating fluid systems naturally produce vortical motions and hence vortices must be important at some level. For example, the Earth's atmosphere, the Earth's oceans, and the atmospheres of the giant planets all produce many different kinds of vortices such as those studied here. In addition, many types of vortices can be easily generated in laboratory experiments (Nezlin & Snezhkin 1993). Thus, one might naively expect vortices to arise naturally in circumstellar disks.

One important constraint on vortex generation is provided by Kelvin's circulation theorem (e.g., Landau & Lifshitz 1987; Ghil & Childress 1987; Shu 1992). This theorem states that for *inviscid barotropic flow*, the number of vortex lines that thread a given area (that moves with the fluid) remains unchanged with time. Thus, for the case of circumstellar disks which are nearly barotropic, vorticity is not automatically "generated" in these systems. In the context of our local equation of motion (see equation [2.20]), the *total* vorticity (per unit surface density) is advectively conserved. This total includes both the vorticity of the perturbation and that of the mean flow. Thus, vorticity can be freely exchanged between the mean flow and the perturbation, but cannot be directly generated (or destroyed). We also note that, except under rather special circumstances, the mean flow is not unstable and does not spontaneously transfer its vorticity into perturbations (vortices).

If the flow is not exactly barotropic, then vorticity can be generated directly. Consider, for example, the force equation [2.16] with a standard pressure term of the form $\rho^{-1}\nabla p$. When we take the curl of the force equation to obtain the equation of motion for the vorticity, we obtain a forcing term of the form

$$\mathbf{F}_V = \frac{1}{\rho^2}\nabla\rho \times \nabla p = \frac{R}{\rho}\nabla\rho \times \nabla T\,, \qquad (4.24)$$

where we have used the ideal gas law $p = \rho RT$ in obtaining the second equality. Clearly, for a barotropic equation of state $p = p(\rho)$, this new term vanishes. However, this forcing term will be nonzero whenever the surfaces of constant temperature do not line up exactly with the surfaces of constant density. Such a situation can occur in a circumstellar disk. In the absence of disk accretion energy, one important heating source for the disk will be reprocessing of stellar photons; this energy source naturally leads to nearly axisymmetric surfaces of constant temperature. On the other hand, gravitational instabilities naturally produce spiral patterns and hence non-axisymmetric surfaces of constant density (e.g., Adams, Ruden, & Shu 1989; Laughlin & Bodenheimer 1994).

We now consider the idealized case in which the surfaces of constant temperature are exactly axisymmetric. We will also assume that the disk has strongly growing spiral density perturbations so that the surfaces of constant density are tilted at an angle $i$ with respect to axisymmetric surfaces. Spiral density wave theory in the WKBJ limit (see Shu 1992; Binney & Tremaine 1987) implies that the tilt angle is given by

$$i \sim \tan i \sim \frac{m}{kr}\,, \qquad (4.25)$$

where $m$ is the azimuthal wavenumber of the perturbation and $k$ is the radial wavenumber. For global modes in circumstellar disks, we expect those with $m = 1$ to be the most important (Adams, Ruden, & Shu 1989; Laughlin & Bodenheimer 1994). We also expect the radial wavelength to be comparable to the radius and hence $i \sim 1/2\pi$. Given this set of approximations, the generalized equation of motion for the vortensity can be written

$$\mathbf{D}\Upsilon \approx \frac{1}{2\pi\sigma} \frac{R}{\rho} |\nabla\rho| |\nabla T| - F_{DIS}, \qquad (4.26)$$

where we have taken $\sin i \sim i \sim 1/2\pi$. For completeness, we have also included a dissipation term $F_{DIS}$.

Next, we want to estimate the timescale $\tau_V$ for vortensity in the disk to be generated by the effect described above. To make this estimate, we write $\mathbf{D}\Upsilon \sim \omega/\sigma\tau_V$. We write the density gradient term as $|\nabla\rho|/\rho \sim \beta/L$, where $\beta$ is the amplitude and $L$ is the sizescale of the spiral perturbation. Finally, we write the temperature gradient term as $|\nabla T| \sim qT/r$, where $q$ is the power-law index of the temperature profile. Solving for the timescale, we obtain the estimate

$$\tau_V \sim \frac{2\pi}{\beta q} \frac{\omega r L}{a^2}, \qquad (4.27)$$

where $a$ is the sound speed. Very roughly, we expect $\beta q \sim 1$, $\omega\Lambda \sim a$, and $L \sim \Lambda$, so that the timescale for the vortensity to change is $\tau_V \sim 2\pi r/a \sim 500$ yr. This timescale corresponds to several orbit times and is thus interesting for the evolution of circumstellar disks. We stress, however, that this argument is extremely crude and must be improved with an honest calculation.

Before leaving this section, we note that several recent papers have begun to address the issue of vortex formation in disks and other rotating systems. For example, the generation of large-scale flow patterns in turbulent astrophysical systems has been studied by Kitchatinov, Rüdiger, & Khomenko (1994); they find that when the rotation rate becomes sufficiently large, the system becomes unstable to the production of large-scale motions which can be considered as vortices. Related work has studied the emergence of coherent vortices out of turbulent flows (McWilliams 1984, 1990).

## 5. ASTROPHYSICAL APPLICATIONS

In this section, we discuss several different astrophysical applications of vortices. The most important applications are in circumstellar disks, where vortices may enhance the formation of giant planets (§5.1) and may provide enhanced energy dissipation for disk accretion (§5.2). For the sake of completeness, we also briefly consider applications of these vortices to the formation of molecular clouds out of the galactic disk (§5.3) and to the formation of substructure within molecular clouds (§5.4).

*5.1 Dust Separation and the Formation of Giant Planets*

One of the main difficulties in forming giant planets (Jupiter, Saturn, Uranus, Neptune) through the mechanism of gravitational instability is that the planets are highly enriched in heavy elements (see, e.g., DeCampli & Cameron 1979; Stevenson 1982; Cameron 1988; Podolak, Hubbard, & Pollack 1993). In the solar nebula from which the planets formed, most of the heavy elements are expected to be in the form of dust grains. For most planet formation scenarios considered previously, the timescale for the dust grains to migrate to the center of the forming protoplanet is too long to explain the observed enrichment. In addition, a solid rocky core is found at the center of the giant planets; it is difficult for such a core to form at all in most previously considered models. As we show below, however, vortices can, in principle, provide a mechanism for separating dust and gas on a shorter timescale and may provide a means of forming giant planets within the gravitational instability scenario.

In a rotating vortex system, the dust grains are not completely coupled to the gas. The grains do not directly feel the pressure forces, but they are coupled indirectly through drag forces. Following convention, we define a response time $\tau_d$ which determines the timescale on which dust grains respond to the force exerted by gas drag. For the regime of parameter space appropriate for circumstellar disks, the response time is given by

$$\tau_d = \frac{\rho_d \langle b \rangle}{2\rho a}, \quad (5.1)$$

where $\langle b \rangle$ is the average radius of the dust grains and $\rho_d$ is the grain mass density (see, e.g., Weidenschilling & Cuzzi 1993). For typical conditions ($\rho_d \sim 2$ g/cm$^3$, $\langle b \rangle \sim 10^{-4}$ cm, $\rho \sim 10^{-12}$ g/cm$^3$, $a \sim 5 \times 10^4$ cm/s), this timescale is relatively short, $\tau_d \sim 2000$ s. Under these conditions, the dust grains quickly reach their terminal velocity, i.e., the force due to gas drag is balanced by the inward radial force (here due to Coriolis effects). This terminal velocity $v_{\text{term}}$ is given by

$$v_{\text{term}} = \tau_d \frac{dh_1}{d\varpi}. \quad (5.2)$$

In order to estimate the timescale for dust grains to settle to the center of the vortex, we use the point vortex solution found analytically in §3. If we use this solution in equation [5.2] and solve the resulting differential equation, we obtain the following simple expression for the timescale $\tau_{\text{settle}}$,

$$\tau_{\text{settle}} = \frac{\pi}{2} \frac{\Lambda^2}{\tau_d \Omega_0 |\Gamma|}, \quad (5.3)$$

where $\Lambda$ is the assumed starting radius of the dust grain and is taken to be the vortex size. To obtain a numerical estimate for the timescale, we let $|\Gamma| \sim a\Lambda$ (corresponding to a weakly nonlinear vortex) and we let the vortex size be the inverse of the Rossby wavenumber, i.e., $\Lambda \sim a/\Omega_0$. We thus obtain the dust settling timescale

$$\tau_{\text{settle}} = 2 \times 10^8 \, \text{yr} \, (\Omega_0 \, 100 \text{yr})^{-2}. \quad (5.4)$$

This timescale is comparable to that expected for dust settling in a hydrostatically supported protoplanet. Furthermore, this timescale is too long to explain the observed element segregation in the giant planets in our solar system. In order to overcome this difficulty, either the grain size must be larger or the vortex strength (as measured by the magnitude of the circulation $|\Gamma|$) must be larger than we assumed here. For example, suppose we require the dust settling timescale to be $\sim 10^6$ yr, roughly two orders of magnitude shorter than given above. This time requirement can be met provided that mean dust radius $\langle b \rangle$ and the circulation $\Gamma$ satisfy the following constraint:

$$\frac{\langle b \rangle}{1\mu\text{m}} \frac{|\Gamma|}{a/k_R} \geq 100 \,. \tag{5.5}$$

Since we naively expect both of the ratios in the above equation to be of order unity, we find that sufficient dust segregation to form giant planets requires somewhat extreme conditions. We note, however, that in principle both the circulation and the grain size *can be* large and hence giant planet formation is at least *possible* is this scenario.

The above simple calculation can be generalized to obtain a more accurate description of dust settling in vortices. Recent papers (Tanga et al. 1995; Barge & Sommeria 1995) have performed numerical integrations of the equations of motion for dust particles in vortices and confirm that dust particles concentrate inside vortices on a relatively short timescale.

Finally, we note that vortices enhance the possibility of forming giant planets via gravitational instability by making the local surface density larger. The perturbation requirements to form a secondary body in a circumstellar disk can be written in fairly general form; this argument is given in Appendix B. The basic result is that a moderately large perturbation, with density contrast $\beta = \Delta\sigma/\sigma \sim$3–5, is required in order to form a secondary body.

*5.2 Energy Dissipation and Disk Accretion*

In this subsection, we briefly discuss how vortices can affect energy dissipation and hence accretion in circumstellar disks. As we discuss below, the overall effect of vortices is to provide additional avenues for energy dissipation in disks. We can identify three conceptually different ways for vortices to participate in energy dissipation: (1) annihilation of vortices, (2) energy dissipation within a single vortex, and (3) a cascade of vortices.

The annihilation of vortices directly leads to the dissipation of energy. The simplest model to consider is that of point vortices with circulation of opposite sign (see §3). The vortices can annihilate, or "short out", in much the same way that magnetic field lines of opposite direction dissipate their energy.

In order to consider the dissipation of energy within a single vortex, we think of the vortex as a small scale analog of the circumstellar disk itself. As shown from our solutions to the equations of motion in the previous sections, the vortices are strongly differentially rotating. Now suppose the fluid has a viscosity $\nu$, which we parameterize in the usual way according to

$$\nu = \frac{2}{3}\alpha a H \sim \alpha a \Lambda \,, \tag{5.6}$$

where $\Lambda$ is the characteristic vortex size and $\alpha$ is the usual dimensionless "$\alpha$ parameter" (see, e.g., Pringle 1981). The viscous diffusion timescale $\tau_D$ is then given by

$$\tau_D \sim \frac{\Lambda^2}{\nu} \sim \frac{\Lambda/a}{\alpha} \sim \frac{1}{\Omega\alpha} \sim \frac{6\,\text{yr}}{\alpha}, \tag{5.7}$$

where we have taken typical values to obtain the numerical estimate. Many studies of viscous disks suggest that $\alpha \sim 10^{-2} - 10^{-4}$ (e.g., Lin & Papaloizou 1980, 1985; Lin 1981), which implies a vortex dissipation timescale of $\tau_D \sim 6 \times 10^2 - 6 \times 10^4$ yr.

Because of the smaller size scale of the vortex, the energy dissipation rate is enhanced over the accretion timescale of the disk as a whole by a factor $f$ given by

$$f \sim (\Lambda/r)^2 \sim 100. \tag{5.8}$$

Thus, for a given fluid viscosity, vortices are more efficient at dissipating energy than the disk as a whole. As a result, vortices can play an important role in energy dissipation and hence disk accretion if a mechanism exists to efficiently transfer energy from the mean flow into the vortex (see §4.5).

Finally, we consider the possibility that these vortices could represent the first stage in a cascade of eddies. We have shown that vortices can arise from a circumstellar disk. We have also shown that the vortices themselves are strongly differentially rotating systems. As a result, smaller vortices can, in principle, arise within the differentially rotating flow of the the larger vortices. These smaller vortices can support still smaller vortices and so forth. In this manner, a cascade of eddies can be set up and can transfer energy from larger scales down to smaller scales. This energy transfer is exactly what is needed to provide an anomalous viscosity in the disk. We stress, however, that we still do not understand the mechanism(s) by which vortices are generated; as a result, this cascade of eddies remains little more than a speculative possibility.

*5.3 Molecular Cloud Formation from the Galactic Disk*

The disk of the galaxy is in many ways a larger version of the circumstellar disks that we have been considering thus far. In particular, vortex motions can arise in the galactic disk and can form substructure within the galaxy (see Petviashvili & Pokhotelov 1992 for a review). The treatment is analogous to that presented above. Although the rotation curve of the galaxy is non-Keplerian, to leading order, only the local rotation speed is important. In any case, the relevant size scale for any structures produced by vortices in the galaxy will be given by the Rossby radius $\Lambda = \sigma_*/\Omega$, where $\Omega$ is the local rotation speed of the galactic flow and where $\sigma_*$ is the velocity dispersion. For typical values for our galaxy, $\Omega \sim 10^{-15}$ rad/s and $\sigma_* \sim 30$ km/s (e.g., Binney & Tremaine 1987), we obtain

$$\Lambda_{\text{galaxy}} = 1\,\text{kpc} \left(\frac{\sigma_*}{30\text{km/s}}\right)\left(\frac{\Omega}{10^{-15}\text{rad/s}}\right)^{-1}. \tag{5.9}$$

Notice that the final size of the structure can be much smaller than this value (after some collapse has taken place). The corresponding mass scale is also fairly large, $M_C \sim \Sigma \Lambda^2$ $\sim 10^7 - 10^8$ $M_\odot$. However, if the cloud formation process is not completely efficient and only about 10% of the mass ends up in the cloud, then this mass scale is about right for giant molecular clouds.

### 5.4 Formation of Molecular Cloud Substructure

It is possible for vortical motions to help produce substructure on the size scale of molecular clouds. The two-dimensional vortices studied in this paper are not well suited to this problem, as molecular clouds are inherently three-dimensional. We note, however, that *any* type of rotating fluid system supports vortical motions and hence vortices *can* occur in these clouds. Very roughly, the relevant size scale for any possible substructure is determined by the Rossby radius $\Lambda = a/\Omega$. Using typical values for the temperature and rotation rates of molecular clouds (e.g., Blitz 1993), we find

$$\Lambda_{\rm cloud} = 1\,{\rm pc} \left(\frac{T}{30{\rm K}}\right)^{1/2} \left(\frac{\Omega}{10^{-14}{\rm rads}^{-1}}\right)^{-1}. \tag{5.10}$$

This size scale is reasonable for molecular cloud substructure. A similar type of self-gravitating vortex has been considered in the context of forming substructure in the interstellar medium (Chantry, Grappin, & Léorat 1993; see also Sasao 1973). In addition, the importance of vorticity in the collapse of molecular clouds and the formation of extended filamentary structures has recently been studied by Monaghan (1994).

## 6. SUMMARY AND DISCUSSION

In this paper, we have begun a study of vortices in circumstellar disks. These vortices are basically storm systems in a generalized geostrophic balance, i.e., a balance between pressure forces, gravitational forces, and the Coriolis force.

### 6.1 Overview of Results

We have obtained a number of basic results concerning the physical properties of vortices in circumstellar disks and other astrophysical systems.

[1] Many different types of vortex solutions are possible in circumstellar disks (see §3–5 and Appendix A). In other words, circumstellar disks can exhibit many different types of vortical motions.

[2] Point vortices are the simplest type of vortex. Their properties can be found analytically (see §3) and hence these vortices provide a prototype for understanding the basic physics of these systems (see also the review of Marcus 1993).

[3] In the absence of self-gravity, the relevant sizescale $\Lambda$ for this type of vortex is roughly determined by the inverse of the Rossby wavenumber, i.e., $\Lambda = a/\Omega$. For circumstellar disks, this sizescale is comparable to the thermal scale height $H$ in the disk.

[4] The leading order effect of self-gravity is to make the effective Rossby wavenumber smaller (§4.1). This effect, in turn, makes the size of the vortex larger than in the case without gravity.

[5] Nonlinear effects can lead to qualitatively new behavior (§4.2). In particular, under special circumstances, solitary wave solutions can arise.

[6] Magnetic fields can, in principle, have two different effects on the evolution of vortices (§4.3). The fields exert a force on the fluid and hence add an additional force term to the equation of motion. For the simplest case of a magnetic field in the vertical ($\hat{z}$) direction, this force corresponds to a magnetic pressure only. The second effect is to allow for ohmic dissipation; however, for expected parameters in circumstellar disks, this effect is small.

[7] Circumstellar disks can, in principle, generate vorticity (or vortensity) through baroclinic effects (§4.5). These effects arise whenever the surfaces of constant density do not line up with the surfaces of constant temperature. Disks can naturally obtain nearly axisymmetric temperature profiles (from reprocessing of stellar photons) and non-axisymmetric density profiles (from self-gravitating spiral modes). The estimated timescale for the vortensity to change is ∼500 yr.

[8] Vortices can enhance the formation of giant planets through the mechanism of gravitational instability (§5.1). In particular, dust grains settle to the center of these vortices. In order for the timescale for dust settling to be interesting for planet formation, either the dust grains must be quite large ($\langle b \rangle \gg 1$ $\mu$m) and/or the vortex circulation must be very strong ($|\Gamma| \gg aH$). Vortices also enhance the formation of giant planets by gravitational instability by increasing the local surface density of the disk.

[9] We have derived a general argument which illustrates the conditions under which secondary bodies can form in circumstellar disks (Appendix B). For moderately massive disks associated with young stellar objects, a moderately large perturbation amplitude $\beta = \Delta\sigma/\sigma \sim 3$–$5$ is required to form a secondary.

[10] Planets formed in vortices should generally rotate in a prograde sense with respect to their orbits (§4.4). In order to produce a planet with retrograde rotation, the circulation of the vortex must be sufficiently strong that the approximations used in this paper break down.

[11] Vortices can enhance the dissipation of energy in circumstellar disks and can thereby help disk accretion (§5.2). For a given fluid viscosity, vortices dissipate energy faster than the disk as a whole. The physical reason for this enhancement is that the differential rotation in the vortex is strong and the size scale is short compared to the entire disk. Thus, vortices can enhance disk accretion, provided that some mechanism exists to transfer energy from the mean circumstellar flow into the vortices.

[12] These vortices can, in principle, also form structure in the galactic disk and within molecular clouds themselves (§5.3 and §5.4). The size of these structures is determined by the local Rossby radius. In the case of the galactic disk, the mass scale of the expected structures is roughly an order of magnitude larger than that of giant

molecular clouds. In the case of rotating molecular clouds, the length scale is ~1 pc, an interesting and relevant sizescale for substructure in molecular clouds.

*6.2 Disk Accretion vs Planet Formation*

In this paper, we have suggested that vortices may be important for both planet formation and for disk accretion. An analogous dual role has been claimed previously for gravitational instabilities in disks (e.g., Adams, Ruden, & Shu 1989), although the secondary bodies are more often considered to be "binary companions" for this latter mechanism. In either case, it might seem paradoxical that one physical process can produce two very different results – namely secondary bodies and disk accretion. However, this apparent contradiction can be resolved for both mechanisms as follows: Energy dissipation and hence disk accretion arises for "failed" structures, whereas secondary bodies can form only within "successful" structures.

We first consider energy dissipation and disk accretion. Self-gravitating spiral modes lead to disk accretion when they saturate at a fairly *low* amplitudes (Laughlin & Bodenheimer 1994), i.e., when the instabilities "fail" to achieve strong nonlinear growth. Similarly, vortices can lead to energy dissipation and hence help disk accretion when they dissipate their energy faster than they can grow, i.e., when they also "fail" to achieve strong nonlinear growth.

On the other hand, secondary bodies can be produced by either mechanism for the case of "successful" structures. Self-gravitating spiral modes can collapse to form secondary bodies when they successfully grow well into the nonlinear regime; as shown in Appendix B, a density contrast $\beta \sim 3-5$ is required (see also Adams & Benz 1992; Laughlin, Bodenheimer, & Yorke 1995). Vortices can lead to giant planet formation if they are sufficiently long-lived (so that dust grains collect at the vortex center) and/or highly nonlinear (so that they must collapse gravitationally), i.e., if the vortices are "successful".

*6.3 Future Work*

This paper represents a preliminary step toward understanding the physics of vortices in circumstellar disks and related astrophysical systems. As we outline below, many directions of future work remain.

The most important issue to be studied is the manner in which vortices are generated in circumstellar disks (see §4.5). This process must be understood in order to determine what types of vortices are produced, how often they are produced, and where in the disk they are produced.

Another important issue is the question of vortex stability and lifetimes. In order for vortices to be interesting for giant planet formation, they must live for many dynamical timescales and hence be relatively stable. On the other hand, vortices which dissipate their energy rapidly will affect disk accretion. It is thus important to determine the conditions under which vortices are long-lived and short-lived.

Many higher order issues also remain unresolved. We have considered only a small fraction of the possible vortex solutions. Vortices can exhibit a wide range of complicated behavior, including vortex merging, vortex scattering, growth and collapse, decay and dispersive spreading, etc. These issues should also be studied.

Finally, we note that the overall goal of this work is to understand the dynamics and evolution of circumstellar disks as a whole. Vortices will play only a partial role in determining the long-term fate of disks. One challenge for the future is thus to integrate vortices into the overall picture of disk evolution.

Acknowledgements

We would like to thank Peter Bodenheimer, Tom Donahue, Curtis Gehman, Frank Shu, and Michael Sterzik for useful discussions. This work was supported by NASA Grant No. NAGW–2802, by an NSF Young Investigator Award, and by funds from the Physics Department at the University of Michigan.

## APPENDIX A: GENERAL SOLUTIONS TO THE EQUATION OF MOTION

We now consider more general solutions to the equation of motion

$$\mathbf{D}\Upsilon = 0 \,, \tag{A1}$$

which implies that the generalized vortensity is advectively conserved. We specialize to the case of steady flow (so that time derivatives are zero) and work in the rest frame of the vortex center; we also neglect the background velocity term $r(\Omega - \Omega_0)$ in the definition [2.21] of the operator $\mathbf{D}$. With these approximations, the equation of motion becomes simply

$$\mathbf{v} \cdot \nabla \Upsilon = 0 \,, \tag{A2}$$

i.e., the velocity of the vortex flow is perpendicular to the gradient of the vortensity $\Upsilon$. Next, we note that the velocity is given by $\mathbf{v} = \hat{z} \times \nabla \Phi$, so that the velocity is also perpendicular to the gradient of the stream function $\Phi$. These two results imply that the gradients of the vortensity $\Upsilon$ and the stream function $\Phi$ must be parallel, i.e.,

$$\nabla \Upsilon = g \nabla \Phi \,, \tag{A3}$$

where $g$ is an unspecified function. This equation can be satisfied provided that the vortensity is a function of the stream function,

$$\Upsilon = \mathcal{G}(\Phi) \,. \tag{A4}$$

To summarize, whenever the generalized vortensity is a function of the stream function (equation [A4]), the equation of motion [A1] is satisfied for conditions of geostrophic flow.

Thus, instead of choosing solutions for the vortensity $\Upsilon$, we can choose a function $\mathcal{G}(\Phi)$ and solve for the stream function from the equation of motion

$$\nabla^2 \Phi - k_R^2 \Phi + \alpha x = \sigma_0 \Upsilon = \sigma_0 \mathcal{G}(\Phi) \,, \tag{A5}$$

which is the appropriate generalization of equation [3.11].

We can generalize the above result further by including the background velocity term $V_B = r(\Omega - \Omega_0)$. With this term, the equation of motion becomes

$$(V_B \frac{\partial}{\partial y} + \mathbf{v} \cdot \nabla)\Upsilon = 0 \,, \tag{A6}$$

where we are again considering the time independent problem. We next define a generalized stream function $\widetilde{\Phi}$ according to

$$\widetilde{\Phi} = \Phi + \int^x V_B \, dx \,. \tag{A7}$$

The same arguments given above then lead to the generalized solution in the form

$$\Upsilon = \mathcal{G}(\widetilde{\Phi}) = \mathcal{G}\left(\Phi + \int^x V_B \, dx\right), \tag{A8}$$

where $\mathcal{G}$ is again an unspecified function.

## APPENDIX B: GENERAL CLUMP ARGUMENT

We now derive a general argument for the size and mass of clumps that form in a self-gravitating circumstellar disk. Suppose that we form a clump in a vortex (such as those considered in this paper) or in the collapsing spiral arm of a disk (as in Adams & Benz 1992). Let the clump mass be $M_C$, the clump size be $\Lambda$, and let $r$ be the radial location of the clump center. This clump must satisfy three general conditions, which we give below.

(1) The clump size must be smaller than the Roche radius of the star/clump system when considered as a binary pair. This condition can be written in the form

$$\Lambda \leq \left(\frac{M_C}{3M_*}\right)^{1/3} r \,, \tag{B1}$$

where $M_*$ is the mass of the star. This expression for the Roche radius is valid in the limit of small clump mass $M_C \ll M_*$; we have also neglected the gravitational potential of the disk itself.

(2) The clump must be gravitationally bound, i.e., the magnitude of the gravitational binding energy must be greater than the thermal energy. Using standard virial theorem results (e.g., Shu 1992), we can write this constraint in the form

$$\Lambda \leq f \frac{G M_C}{a^2} \,, \tag{B2}$$

where we have used an isothermal equation of state and where the quantity $f$ is a dimensionless number. For a clump with a modest central concentration, we expect $f \sim 1/10$.

(3) Finally, we require that the perturbation which produced the clump was produced by the background circumstellar disk. This requirement can be written in terms of the amplitude of the perturbation and takes the form

$$M_C = \beta \sigma_0(r) \pi \Lambda^2 , \tag{B3}$$

where the parameter $\beta \equiv \sigma/\sigma_0$ measures the amplitude of the perturbation.

We can combine the above three constraints to obtain the allowed range of clump sizes or masses forming in a circumstellar disk. We find that the allowed mass range is given by

$$\frac{\beta^3 \sigma_0^3 \pi^3 r^6}{9 M_*^2} \geq M_C \geq \frac{a^4}{\beta \pi \sigma_0 f^2 G^2} . \tag{B4}$$

As expected, for some range of parameters, the above range of allowed clump masses ceases to exist. Also as expected, the requirement for real clump sizes can be written in terms of the Toomre parameter $Q_T$ as follows:

$$Q_T < \beta (f/3)^{1/2} \frac{\kappa}{\Omega_*} , \tag{B5}$$

where $\kappa$ is the epicyclic frequency and $\Omega_*$ is rotation rate due to the stellar potential only, i.e., $\Omega_*^2 = GM_*/r^3$. For the expected rotation curves in circumstellar disks, $\kappa \sim 2\Omega_*$. Using representative values for $f$ and $\kappa/\Omega_*$, we can write our result in terms of the required perturbation amplitude for clump formation:

$$\beta > \frac{5}{2} Q_T . \tag{B6}$$

Thus, we see that fairly large perturbation amplitudes are required for clump formation.

This finding is consistent with the results obtained from numerical simulations of self-gravitating disks (e.g., Laughlin & Bodenheimer 1994). These simulations show that when disks are near stability, gravitational instabilities tend to saturate at somewhat low amplitudes (perturbations of order unity) and disk accretion occurs. On the other hand, for disks which are more unstable, the gravitational instabilities grow to larger amplitudes and the fragmentation of the disk into clumps can occur (see also Bonnell & Bate 1994).

## FIGURE CAPTIONS

Figure 1. Schematic diagram of local coordinate system.

Figure 2. Schematic diagram illustrating geostrophic balance. For the case shown, both the coriolis force ($C$) and the gravitational force ($g$) point radially inward, whereas the pressure force ($P$) is directed radially outward.

Figure 3. Surface density perturbations for linear vortices (surface density is given in arbitrary units). For these solutions, the vortensity profiles are step functions in local radius. The radial size of the area of nonzero vortensity is $R$. Profiles are shown for $k_R R = 1/5$ and 5.

Figure 4. Velocity fields for linear vortices with step function vortensity profiles. Notice that the flow is clockwise in the rotating frame of reference moving with the vortex center. Cases are shown for (a) $k_R R = 1/5$ and for (b) $k_R R = 5$.

Figure 5. Diagram showing difference between adverse and prograde vortices. When the vortical motion of the flow opposes the background shear flow (a), the adverse vortex gets ripped apart and has a short lifetime. In the opposite case (b) in which the vortical motion goes with the background shear flow, the prograde vortex can be long-lived.